\documentclass[aps,physrev,twocolumn,groupedaddress]{revtex4-2}

\usepackage{graphicx}
\usepackage{amsmath}
\usepackage{lineno}
\usepackage{gensymb}

\begin{document}


\title{Ab initio investigation of electronic and lattice properties of Fe$_4$(P$_2$O$_7$)$_3$}


\author{Svitlana Pastukh$^a$}
\author{Pawe\l{} T. Jochym$^a$}%

\author{Jan \L{}a\.zewski$^a$}%
\author{Dominik Legut$^{b,c}$}%
 \email{Contact author: dominik.legut@vsb.cz}
\author{Przemys\l{}aw Piekarz$^a$}%
\affiliation{%
 $^a$Institute of Nuclear Physics Polish Academy of Science, PL-31342 Kraków, Poland\\
 $^b$Department of Condensed Matter Physics, Faculty of Mathematics and Physics, Charles University CZ-121 16, Prague, Czech Republic\\
 $^c$IT4Innovations, VSB-Technical University of Ostrava CZ-708 00, Ostrava, Czech Republic
}%


\begin{abstract}
 In this research, we examine the electronic, magnetic, and lattice properties of the Fe$_4$(P$_2$O$_7$)$_3$ compound using the first principles calculations based on the density functional theory. The crystal lattice has a monoclinic structure, belonging to the P$2_1/n$ space group. The optimized lattice parameters are a=7.406 \AA, b=21.425 \AA, c=9.529 \AA, and agree very well with the experimental data, thanks to the local Coulomb interactions and van der Waals forces included in the calculations. The investigation considers several magnetic orderings. The lowest total energy was found for the antiferromagnetic configuration with the magnetic moment of $\sim4.6~\mu_{\text{B}}$ per Fe atom. The electronic structure calculation shows the Mott insulating state with the energy gap $E_g=2.87$~eV. For the relaxed crystal structure, the elastic properties were obtained and analyzed. The phonon dispersion relations and density of states were calculated within the temperature-dependent effective potential methodusing atomic multidisplacements obtained by high efficiency configuration space sampling.
\end{abstract}


\maketitle


\section{Introduction}
The iron pyrophosphate compounds are of great scientific and technological interest due to their diverse chemical and physical properties and a broad range of applications.
Iron pyrophosphate shows high chemical stability in a solution and can be applied as an active sorbent for the removal of radionuclides from water \cite{liu2014facile}. The recent study revealed that iron pyrophosphate doped with carbon can be successfully used to control tetracycline antibiotic pollution in aqueous solution by peroxymonosulfate activation \cite{ma2022iron}. An enhanced photoelectrochemical water oxidation activity of nanocrystalline iron pyrophosphate in amorphous iron phosphate overlayer was also reported \cite{xia2022nanocrystalline}. A new compound of iron pyrophosphate with added cobalt carbonate was shown to be prospective for environmental applications \cite{boonchom2009synthesis}. These compounds are also intensively studied for biomedical applications. Numerous studies indicate the important role of ferric pyrophosphate in iron deficiency treatment and iron transport control in human body \cite{gupta1999dialysate, fidler2004micronised, trivedi2021delivery}. 
In addition, iron pyrophosphate has attracted attention due to its potential applications in electrochemical energy storage devices, such as lithium-ion batteries. Porous iron pyrophosphate nanostructure was shown to possess an enhanced electrocatalytic activity in Li-O$_2$ batteries \cite{lee2020redox}, while lithium iron pyrophosphate compound synthesized by conventional solid-state reaction was reported as promising cathode for a lithium-ion battery system for large-scale applications \cite{nishimura2010new}.

Although the above studies demonstrate a broad range of iron pyrophosphate applications, there is a limited number of investigations concerning its crystal structure, electronic, and magnetic properties. There are four types of such compounds reported in the literature:  Fe$_2$P$_2$O$_7$, Fe$_3$(P$_2$O$_7$)$_2$, Fe$_4$(P$_2$O$_7$)$_3$ and Fe$_7$(P$_2$O$_7$)$_4$ \cite{hoggins1983crystal, xiao2007solid, padhi1997effect, elbouaanani2002crystal}, varying by the crystal structure and the way the octahedral iron ions link to each other. In particular, to the best of our knowledge, there are no {\it ab initio}, theoretical studies of electronic and lattice dynamical properties of these materials.

In the current study, we focus on the Fe$_4$(P$_2$O$_7$)$_3$ compound with the monoclinic P2$_1/n$ symmetry. The structural description and chemical characterization of this material was done by Pahdi et al. \cite{padhi1997effect}, who reported presence of two distinct Fe$_2$O$_9$ groups of face shared octahedra and three kinds of the P$_2$O$_7$ groups. More detailed analysis of the crystal structure and magnetic properties of Fe$_4$(P$_2$O$_7$)$_3$ was done by Elbouaanani et al. \cite{elbouaanani2002crystal}, using neutron diffraction and M\"{o}ssbauer spectrometry techniques. The authors reported the existence of four antiferromagnetic iron sublattices, which correspond to the four crystallographically distinct positions of iron atoms. Also, the existence of a paramagnetic to antiferromagnetic transition was observed at about $T = 50$~K.  This phase transition, as a result of Fe$^{3+}$ magnetic ordering, was also confirmed in the low temperature heat capacity study \cite{shi2013low}.
In general, iron pyrophosphate research covers a wide range of experimental topics such as synthesis methods, analysis of structure and magnetic properties, and exploration of catalytic applications. However, there is currently a lack of {\it ab initio} studies that delve into the electronic band structure and the impact of local Coulomb interactions. Furthermore, the lattice dynamical properties of these compounds remain unexplored. 

Here, we investigate the structural, electronic, and dynamical properties of Fe$_4$(P$_2$O$_7$)$_3$ using density functional theory (DFT), with the local Coulomb and van der Waals (vdW) interactions included within the DFT+U method.
The obtained lattice parameters show good agreement with the experimental data.
Also, the magnetic configuration for the ground state corresponds well to the results of neutron measurements.
The calculated phonon dispersion relations provide further insight into the dynamical properties of the material.


\section{Calculation method}
The Vienna Ab initio Simulation Package (VASP)~\cite{Kresse, KresseGF} with the projector augmented-wave method~\cite{Blochl1994} and the generalized gradient approximation~\cite{PBE} was used for relaxation of the crystal structure and to calculate the force constants. The geometry optimization was carried out within a unit cell, which is equivalent to the primitive cell, containing a total of 124 atoms. 
In order to determine the magnetic ground state, three different antiferromagnetic (AFM) arrangements were considered.
To effectively sample the Brillouin zone, integration in the reciprocal space was performed over $4\times2\times4$ k-point mesh generated with the Monkhorst-Pack scheme~\cite{MP}. To ensure high precision of the calculations, a cut-off energy of 500 eV was applied.

Computations were carried out using the DFT+U technique, incorporating distinct parameters for iron and oxygen atoms. The primary objective of these calculations was to gain insights into the electronic properties of iron pyrophosphate. The results presented here were obtained using $U_d$=8~eV and $J_d$=0.9~eV for the Fe($3d$) states and $U_p$=4~eV and $J_p$=0.5~eV for the O($2p$) states. This parameter set resulted in an insulating state with an electronic band gap $E_\text{g}$=2.87~eV. For this specific combination of $U$ and $J$ parameters, we obtained a good agreement with the experimental lattice parameters, as shown in Tab.~\ref{tabfe}.
It should be noted that calculations performed with smaller Coulomb parameters ($U_d=4$~eV, $U_p=2$~eV) as well as without
the local interactions ($U_d=U_p=J_d=J_p=0$) did not converge to a well-defined ground state. This indicates the significance of local Coulomb interactions in iron pyrophosphate.

To overcome the limitations of conventional approximations in accurately representing the long-range dispersive interactions during crystal structure optimization, we included approximate van der Waals (vdW) forces into the process. This involved incorporating vdW corrections proposed by Grimme, namely the D2 \cite{D2} and D3 \cite{D3} methods, as well as those suggested by Tkatchenko and Scheffler (TS) \cite{TS1}. Our objective in incorporating these corrections was to refine the treatment of interatomic interactions and elevate the precision of crystal structure optimization.

The mechanical stability of Fe$_4$(P$_2$O$_7$)$_3$ was investigated by computing its elastic constants at $T = 0$~K using the AELAS code \cite{zhang2017aelas}.The mechanical properties, such as Young's modulus, shear modulus, elastic anisotropy, as well as Poisson's and Pugh's ratios, were also derived and analyzed.

Phonon dispersion relations and phonon density of states (PDOS) were computed utilizing the temperature dependent effective potential (TDEP) approach~\cite{Hellman2011,Hellman2013} and the Alamode software~\cite{Tadano2014}. To obtain the force-constant matrices, the atomic multidisplacements were generated by the high efficiency configuration space sampling (HECSS)~\cite{Jochym2021}, and followed by the calculation of the resulting Hellmann-Feynman forces acting on all atoms. 
Phonon energies and polarization vectors were subsequently determined through the exact diagonalization of the dynamical matrix. The calculations were performed for the $2\times1\times1$ supercell containing 248 atoms. Furthermore, the LO-TO splitting was calculated using the static dielectric tensor and Born effective charges obtained within density functional perturbation theory \cite{Gajdos2006}.

\section{Crystal structure and magnetic order}
\begin{figure*}[t!]
\centering
   \includegraphics[width=200mm]{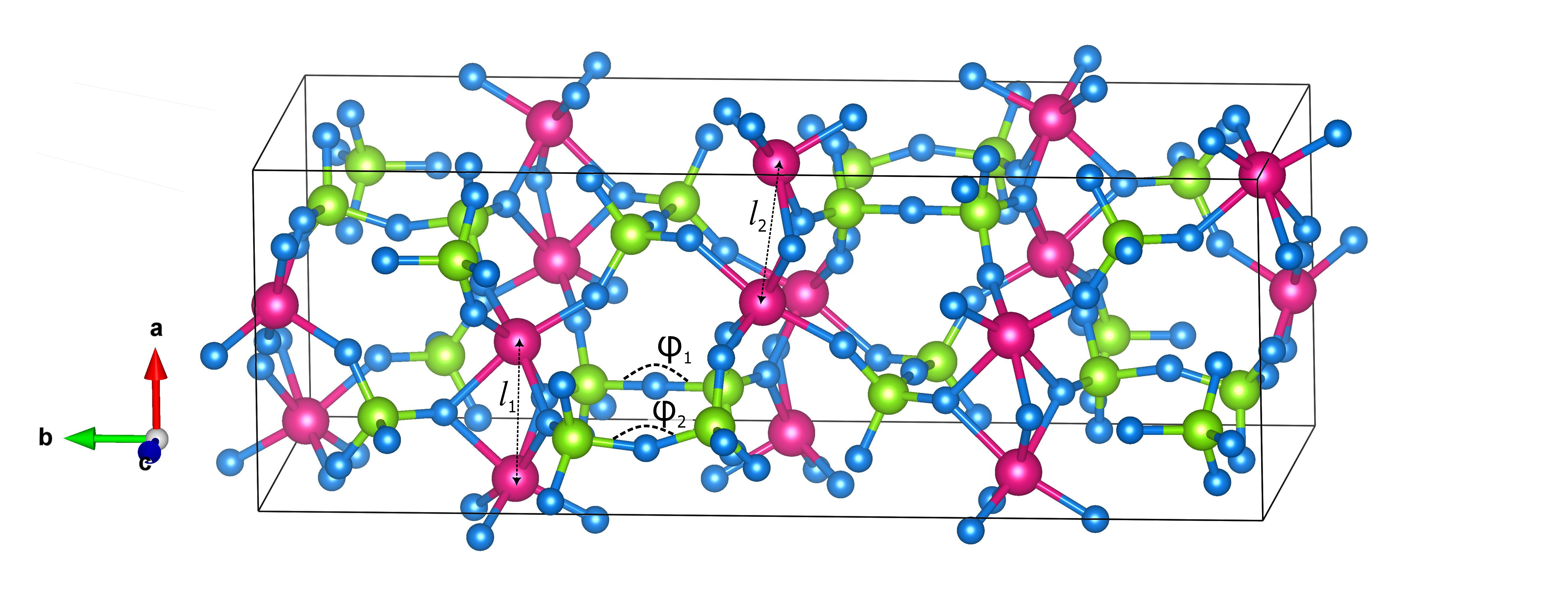}
   \caption{Crystal structure of Fe$_4$(P$_2$O$_7$)$_3$. Iron, phosphorus, and oxygen ions are represented as pink, green, and blue balls, respectively. The lengths $l_1$, $l_2$ and angles $\varphi_1$, $\varphi_2$ denote the intra-dimer Fe-Fe distances and bridging bond P-O-P angles, respectively. The image was rendered using VESTA software~\cite{momma2011vesta}.}
   \label{fe1}
\end{figure*}

At room temperature, Fe$_4$(P$_2$O$_7$)$_3$ crystallizes in the monoclinic structure with the $P2_1/n$ (14) space group and its primitive unit cell contains four chemical formula units \cite{elbouaanani2002crystal}. There are four different iron crystallographic sites surrounded by deformed octahedra. Six distinct diphosphate groups (P$_2$O$_7$) provide six terminal oxygen atoms, which act as their coordination ligands. As depicted in Fig.~\ref{fe1}, two types of octahedra joined by three common oxygen atoms (sharing one face) create double octahedra.
This arrangement causes significant shortening of distances between iron atoms in two different Fe$_2$O$_9$ dimers, as reported in 
Ref.~\cite{elbouaanani2002crystal}: \textit{l}$_1$=2.967~\AA, and \textit{l}$_2$=3.048~\AA. These experimental values were very well reproduced with the DFT calculations, which yielded \textit{l}$_1$=3.005~\AA, and \textit{l}$_2$=3.048~\AA.
The atomic configuration also includes three non-equivalent types of diphosphate groups. In each case, the coordination of iron atoms involves all six oxygen atoms, except for the bridging oxygen atom, which remains unengaged. It results from the specific geometric arrangement where a mirror plane, perpendicular to the lattice's $a$-axis, includes both phosphorus atoms, the bridging oxygen atom, and one of the terminal oxygen atoms. It's important to note that in this structure, two variations of P$_2$O$_7$ anions exist with two different bridging bond angles \text{$\varphi$}$_1$=179.81$\degree$ and \text{$\varphi$}$_2$=148.20$\degree$.

In our investigation, we performed calculations for the ferromagnetic (FM) and three antiferromagnetic (AFM) states. The results indicate that the energy of the FM state is higher than that of the AFM states, with an energy difference of $\Delta E = 0.135$~eV per formula unit. Consequently, we focused our analysis on the AFM configuration.

\begin{figure}[t!]
\centering
  \includegraphics[width=90mm]{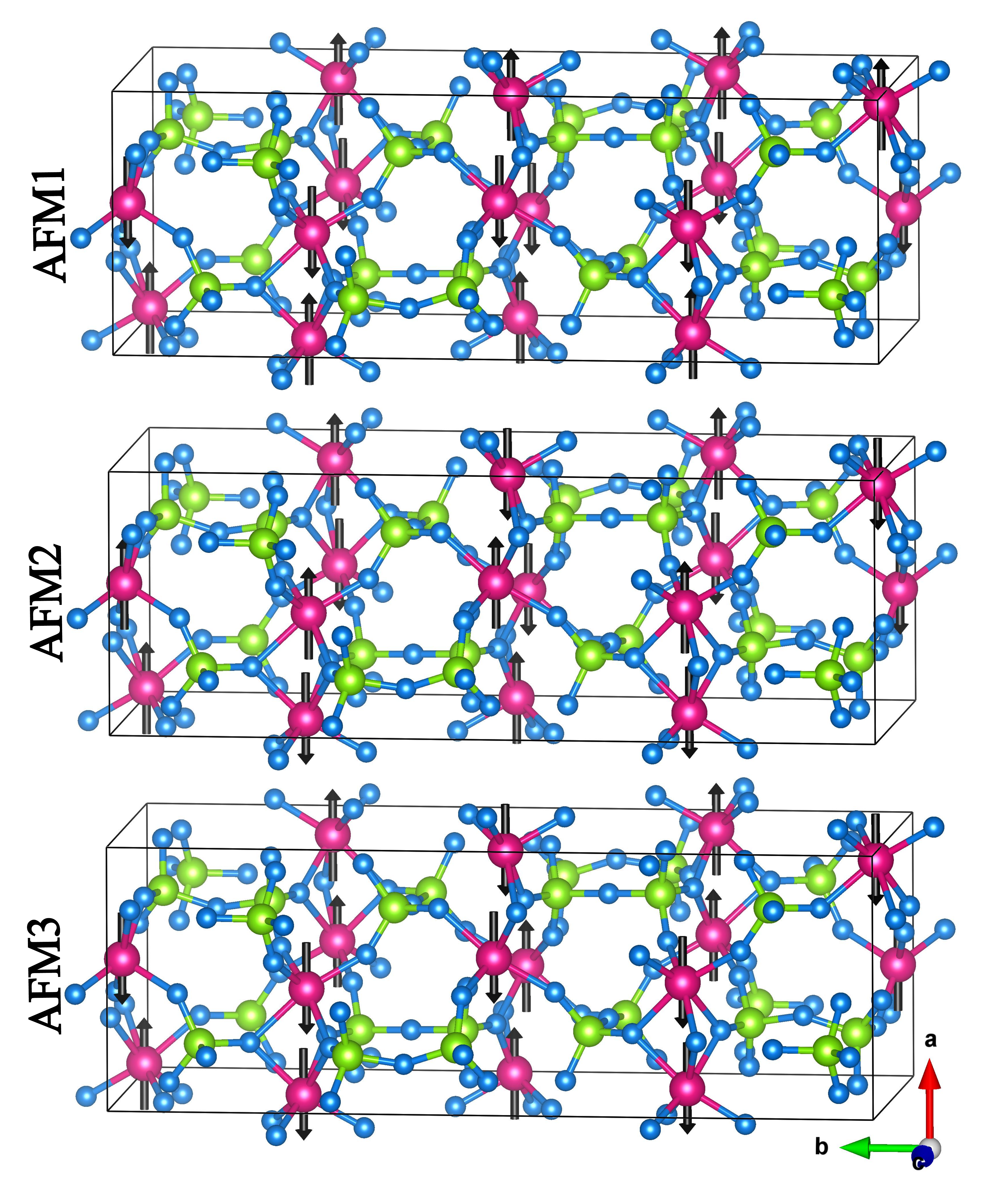}
\caption{Representation of magnetic moments on the iron atoms of Fe$_4$(P$_2$O$_7$)$_3$ for three AFM configurations.}
  \label{AFMfe}
\end{figure}

For the three different AFM configurations (AFM1, AFM2, and AFM3) shown in Fig.~\ref{AFMfe}, the cell parameters and magnetic moments on Fe atoms were computed and compared with the experimental data in Tab.~\ref{tabfe}.
\begin{table}[b!]
    \centering
    \caption{Lattice parameters and magnetic moments in iron pyrophosphate compared with the experimental data.}
    \begin{tabular}{c|c|c|c|c}
    \hline \hline
         &AFM1&AFM2&AFM3& Experimental \cite{elbouaanani2002crystal}\\ \hline
         $a$ (\AA)& 7.441 &7.441&7.438& 7.389\\
         $b$ (\AA)& 21.571 &21.569&21.571& 21.337\\
         $c$ (\AA)&  9.583 &9.583&9.585& 9.517\\ 
         $\beta$ (deg)&90.07&90.08&91.27&90.03\\ \hline
         $\Delta E$ (eV)& 0 & 0.067 &0.07&--\\
         $m$ ($\mu_B$)&  4.6 &4.6 &4.6& 4.55(5)\\
    \hline\hline
    \end{tabular}
    \label{tabfe}
\end{table}
The AFM1 pattern reflects the magnetic configuration observed via neutron diffraction measurements below the N\'{e}el temperature $T_\text{N}=50$~K~\cite{elbouaanani2002crystal}. This arrangement shows reverse magnetic moments on two Fe$^{3+}$ ions within the dimers (along the $a$ direction), while nearest-neighbor iron atoms along the $c$ direction exhibit the ferromagnetic (FM) order. The antiparallel orientation of moments within the Fe dimers stems from a direct AFM interaction due to relatively short Fe-Fe distance and indirect superexchange interactions along the Fe-O-Fe path~\cite{elbouaanani2002crystal}. Interactions between dimers reveal a frustrated nature due to both FM and AFM superexchange pathways via phosphorus ions.
AFM2 displays moments in opposite directions along the $a$ and $c$ axes, while AFM3 features moments arranged in antiparallel alignment along the $c$ axis and parallel magnetization within the dimers. The computed magnetic moment ($m=4.6~\mu_{\text{B}}$/Fe atom) and lattice parameters exhibit negligible sensitivity to the specific AFM configuration. Only the monoclinic angle $\beta$ for AFM3 slightly deviates from the experimental value (refer to Tab.~\ref{tabfe}). This configuration also possesses the highest total energy due to the parallel alignment of magnetic moments within the Fe dimers. For each AFM configuration, the magnetic moments at distinct Fe crystallographic sites are slightly different, and their values are presented in Supplementary Material (SM).

The identified ground state with the AFM1 configuration closely corresponds to experimental outcomes, displaying a very good agreement with the measured magnetic moment of $m=4.55~\mu_\text{B}$ at $T=1.7$ K~\cite{elbouaanani2002crystal}. 
Additionally, inclusion of the vdW attractive forces reduces the lattice constants, significantly improving their values (refer to Tab.~\ref{tabfe2}). 
The three different types of vdW corrections give similar lattice parameters, and the best agreement with the experimental data was achieved for the D2 method. In the following investigations, we focus on exploring the electronic and phonon properties exclusively for the AFM1 ground state configuration.

\begin{table}[h!]
    \centering
    \caption{Lattice parameters of iron pyrophosphate computed for three types of vdW corrections and compared with the experimental data.}
    \begin{tabular}{c|c|c|c|c}
    \hline \hline
         &D2&D3&TS& Experimental \cite{elbouaanani2002crystal}\\ \hline
         $a$ (\AA)& 7.406 &7.432&7.403& 7.389\\
         $b$ (\AA)& 21.425 &21.521&21.492& 21.337\\
         $c$ (\AA)&  9.529 &9.556 &9.522& 9.517\\ 
         $\beta$ (deg)&90.07&90.08&91.27&90.03\\ \hline
    \hline
    \end{tabular}

    \label{tabfe2}
\end{table}

\section{Electronic properties}
\begin{figure}[b!]
\centering
   \includegraphics[width=85mm]{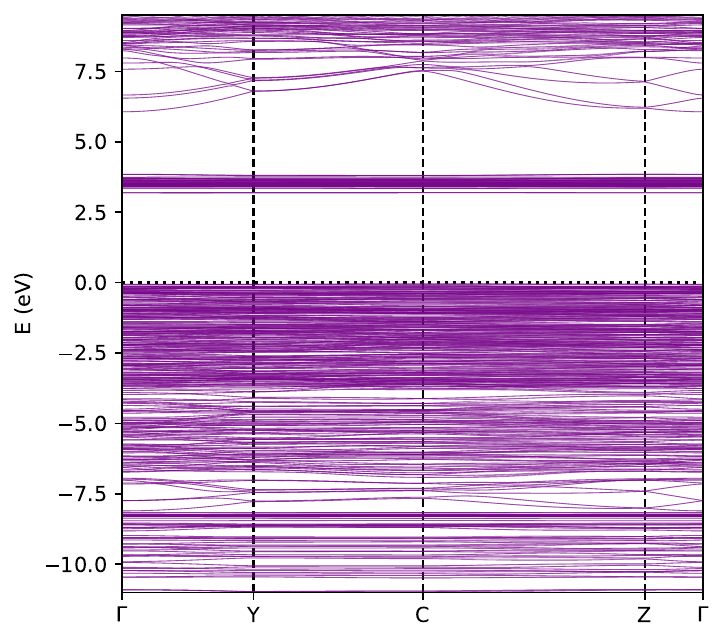}
   \caption{Electronic band structure of iron pyrophosphate. The Fermi level (dotted line) was set to zero.}
   \label{bsfe}
\end{figure}

\begin{figure}[t!]
\centering
   \includegraphics[width=85mm]{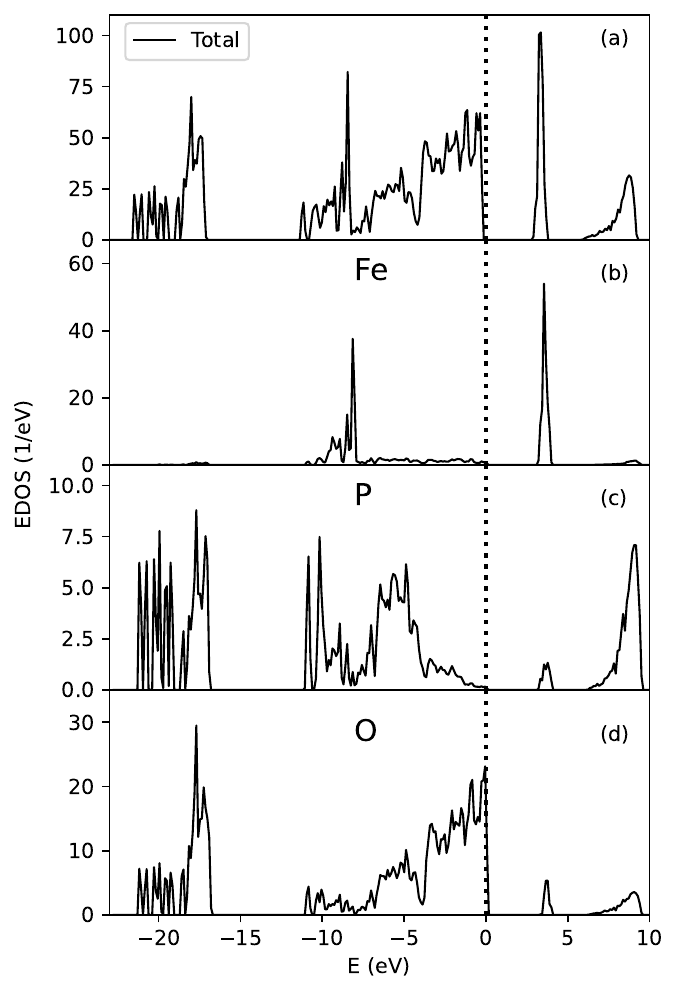}
   \caption{The total and partial electronic density of states of iron pyrophosphate, with the Fermi level (dotted line) set at zero.}
   \label{Dosfe}
\end{figure}

The electronic band structure of Fe$_4($P$_2$O$_7$)$_3$ shows the Mott insulating state with the band gap $E_\text{g}=2.87$~eV (see Fig. \ref{bsfe}), which separates a narrow, flat conducting band from a wide valence band spanning approximately 11~eV, and very weakly depends on the wave vector.
The AFM1 ordering results in the same total electronic density of states (EDOS) for both spin components (refer to Fig.~\ref{Dosfe}). An examination of the partial EDOS, which showcases the electronic states projected onto various iron atom types, reveals the presence of exchange splitting within the $3d$ states (see the EDOS for the four distinct iron atoms shown in Fig. [S2] in SM). Within the energy range spanning from -7 eV to the Fermi energy ($E_\text{F}$), weak contributions from both spin components are evident.
A main band, centered roughly at -8 eV, primarily consists of majority-spin states, constituting the lower Hubbard band. The unoccupied minority-spin $3d$ states displays a notable peak at around 4 eV, forming the upper Hubbard band.
Within the energy range spanning from -12 eV to $E_\text{F}$, the Fe($3d$) states demonstrate hybridization with the O($2p$) states, with the latter prevailing in the vicinity of $E_\text{F}$. The O($2s$) states are localized within the energy range between -22 and -17~eV. In the empty bands situated above the forbidden gap, a minor contribution stemming from oxygen $s$ and $p$ states becomes discernible. Additionally, the $3p$ states inherent to the phosphorus atoms remain occupied within the energy span from -11 eV to $E_\text{F}$, accompanied by an extra peak discernible around -11~eV. The slender bands below -17~eV primarily accommodate the P($3s$) states.


\section{Mechanical stability}

In order to confirm the intrinsic phase stability of the AFM1 phase, we determine the elastic constants ($C_{ij}$) at $T=0$~K using the AELAS code \cite{zhang2017aelas}. 
For the monoclinic and lower symmetry 
crystals, one usually inspects the eigenvalues of the
elastic constants tensor or the corresponding compliance
coefficients tensor. The crystal phase with the monoclinic
symmetry is described by thirteen independent elastic
constants, see  Ref. \cite{legut2019} and references therein. 
The $C_{ij}$ values for the AFM1 phase are listed in Table \ref{Cij}.

\begin{table}[h!]
    \centering
    \caption{The elastic constants ($C_{ij}$) in GPa for the AFM1 phase of Fe$_4($P$_2$O$_7$)$_3$.}
    \begin{tabular}{ccccccc}
    \hline
$C_{11}$ & $C_{12}$ & $C_{13}$ & $C_{15}$ & $C_{22}$ & $C_{23}$&$C_{25}$ \\
168.51 &  43.80 & 58.00 & -6.84 & 162.87  & 53.47 & -8.41 \\   
\hline
$C_{33}$ & $C_{35}$ & $C_{44}$ & $C_{46}$ & $C_{55}$ & $C_{66}$ \\
             \hline
         179.01 & -5.93 & 52.18 & -0.84 & 52.74 & 18.90 \\
    \hline
    \end{tabular}
    \label{Cij}
\end{table}

Some of the $C_{ij}$ are negative, however, this is allowed for the monoclinic symmetry. The important fact is that all eigenvalues of the elastic constants tensor are positive, and therefore this compound is intrinsically stable in monoclinic structure. Using the single crystal $C_{ij}$ values, we further elaborate on other mechanical characteristics of Fe$_4($P$_2$O$_7$)$_3$. The polycrystaline elastic behavior was derived using the Hill averaging scheme \cite{Hill_1952} which combines the Reuss and Voigt\cite{Grimvall} limits of bulk ($B_R$,$B_V$) and shear ($G_R$,$G_V$) moduli. This allows us to determine the averaged bulk, shear, and Young's moduli:
\noindent
\begin{equation*}
E=\frac{9B_{RVH}G_{RVH}}{3B_{RVH}+G_{RVH}}
\end{equation*}

\noindent as well as the Poisson ratio:
\begin{equation*}
\nu=\frac{3B_{RVH}-E_{RVH}}{6B_{RVH})},
\end{equation*}

\noindent and elastic anisotropy:
\begin{equation*}
A_{RVH}=\frac{(G_V-G_R)}{(G_V+G_R)}
\end{equation*}
see the table \ref{elast}. 

\begin{table}[h!]
\centering
\caption{The Hill averaged values of bulk ($B_{RVH}$), shear modulus ($G_{RVH}$), elastic anisotropy ($A_{RVH}$), the Pugh ratio ($B_{RVH}/(G_{RVH}$), Young modulus ($E$), and the Poisson ratio ($\nu$). The units for ($B_{RVH}$), ($G_{RVH}$), ($E_{RVH}$) is GPA, the $A_{RVH}$, Pugh and Poisson ratios are dimensionless.}
\begin{tabular}{cccccc}
\hline
$B_{RVH}$ & $G_{RVH}$ & $A_{RVH}$ & $B_{RVH}/G_{RVH}$ & $E_{RVH}$ & $\nu$ \\
\hline
90.5 &  44.2 & 0.096 & 2.05 & 113.9 &  0.29\\
\hline

\end{tabular}
\label{elast}
\end{table}

The elastic anisotropy  within the interval (0–1) indicates the deviation from the elastically isotropic behavior, {\it i.e.}, $A=0$, which indeed is the case here, as $A=0.096$ is much lower than similar low-dimensional frustrated antiferromagnets like KCuF$_3$ and Cu(en)(H$_2$O)$_2$SO$_4$, where A=0.26 and A=0.20, respectively \cite{legut2019}. To assess the ductility or brittleness of Fe$_4($P$_2$O$_7$)$_3$ we use the values of $B_{RVH}$, $G_{RVH}$, $E_{RVH}$,and $\nu$. In contrast to KCuF$_3$ and Cu(en)(H$_2$O)$_2$SO$_4$, Fe$_4($P$_2$O$_7$)$_3$ is definitively not soft, as $B_{RVH}$ is rather moderate. The Pugh ratio ($B_{RVH}/G_{RVH}$) is over factor 2 indicating ductility of the compound similar to the above-mentioned AFM compounds. Similarly, the ductility is correlated with the lower values of $E_{RVH}$ and $G_{RVH}$ as well as having
the Poisson ratio ($\nu$) below 0.5, indicating weaker and less directional bonding, {\it i.e.} more isotropic behavior \cite{CHU1997147}. The directional Young moduli are obtained from the compliance tensor elements as 
$E_x=S^{-1}_{11}=145$ GPa, $E_y=S^{-1}_{22}=142$ GPa, and $E_z=S^{-1}_{33}=149$ GPa, again indicating very small elastic anisotropy despite large difference in length of the lattice vectors. Let us note, that these directional Young moduli are higher than the polycrystalline averaged one E due to the monoclinic structure. 

Inserting the density, $\rho=3.219g/cm^{-3}$ from our calculated volume and the averaged $B_{RVH}$, $G_{RVH}$ one determines the longitudinal $v_l=\sqrt{B_{RVH}+(4/3)G_{RVH}/\rho}$ and transverse $v_t=\sqrt{G_{RVH}/\rho}$ sound velocities in this compound\cite{ANDERSON1963909}. The calculated values amount to $v_l=2154.6$ m.s$^{-1}$ and $v_t=1171.8$ m.s$^{-1}$, respectively.
\section{Dynamical properties}
In this section, we delve into the lattice dynamical properties of iron pyrophosphate. 
The primitive cell contains 124 atoms, which results in a 372 phonon dispersion relations. The phonon dispersion relations and density of states obtained for the primitive cell are presented in SM. In this case, most frequencies are real, except for four phonon modes near the Y and C points. Notably, the lowest soft mode in the primitive cell is identified at the Y point with the wave vector $q=(0.5,0,0)$. Therefore, we doubled the crystal structure along the $a$ direction to investigate the dynamical properties and stability of this material.

\begin{figure*}[tp!]
\centering
   \includegraphics[width=0.98\textwidth]{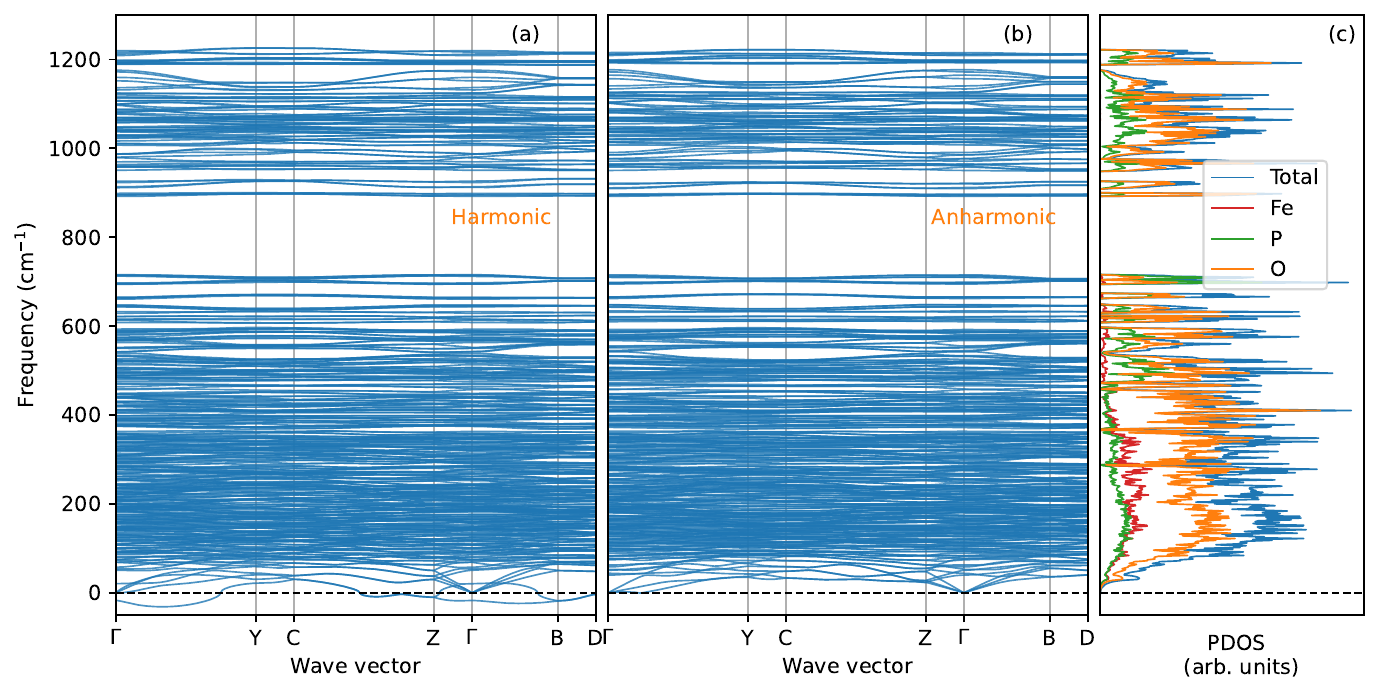}
   \caption{The phonon dispersion relations calculated using a) harmonic approximation and b) third-order anharmonic correction,  and c) the total and partial element-projected PDOS for anharmonic calculation in Fe$_4$(P$_2$O$_7$)$_3$.}
   \label{Phfe}
\end{figure*}

Fig.~\ref{Phfe}(a, b) presents the phonon dispersion curves calculated for the $2\times1\times1$ supercell obtained within the harmonic approximation and with the third-order anharmonic correction included in the atomic potential at $T=100$~K.
Comparing to the result obtained for the primitive cell (see SM), we observe the stabilization of 
the soft modes at the Y and C points, and only weak soft modes close to the $\Gamma$ and B points are found in the harmonic calculations. However, the third-order anharmonic corrections remove the imaginary modes, showing the significant influence of anharmonicity on dynamical properties. Therefore, the inclusion of anharmonic corrections significantly impacts the overall stability and allows for a more accurate representation of the atomic vibrations and interactions within the structure.
This result agrees with the elastic data analysis, confirming the mechanical stability of the system.

Fig.~\ref{Phfe}(c) displays the total phonon density of states (PDOS) along with the partial contributions from Fe, O, and P vibrations calculated with the third-order anharmonic correction. In the low-frequency range (0–400 cm$^{-1}$), O vibrations predominantly govern phonon states, with comparatively smaller contributions from Fe and P atoms. Between 400-700 cm$^{-1}$, oxygen atom vibrations dominate, while around 720 cm$^{-1}$, phosphorous atom vibrations prevail. At higher frequencies (approximately 900-1250 cm$^{-1}$), an elevated density of O atom vibrations is noticeable, with slightly lesser contributions from P atom vibrations.


\section{Summary}
In summary, the presented study of the structural, electronic, and dynamical properties of Fe$_4$(P$_2$O$_7$)$_3$ has yielded new knowledge about this compound. The calculations carried out for three different antiferromagnetic orderings allowed to determine the ground state configuration, which closely match the experimental data. By using the DFT+U approach, the Mott insulating state with the electronic band gap $E_\text{g} = 2.87$~eV was predicted.

The examination of the crystal structure demonstrated the existence of coordinated terminal oxygen atoms from diphosphate groups, resulting an interconnected double octahedron of iron ions. The accuracy of the computational approach was validated by the calculated lattice parameters, iron-iron separations, and bond angles, which showed good agreement with the experimental data. Additionally, the analysis revealed that Fe$_4$(P$_2$O$_7$)$_3$ contains three different types of crystallographically non-equivalent diphosphate groups, each of which adds to the overall structural stability of the compound.
The analysis of the obtained elastic parameters confirmed the mechanical stability of the crystal structure.

The lattice dynamics of iron pyrophosphate was investigated by calculating phonon dispersion relations and phonon density of states using the harmonic and third-order anharmonic approximations within the TDEP approach. The analysis of the phonon dispersion curves identified soft modes; however, these were stabilized by taking into account anharmonicity of the lattice.
\section{Acknowledgement}
D.L. acknowledges financial support by the projects No. 22-35410K from the Czech Science Foundation and 
No. CZ.02.01.01/00/22\_008/0004572 by the Ministry of Education, Youth, and Sports of Czechia.  

\bibliography{apssamp}

\end{document}